\documentstyle[preprint,prd,aps]{revtex}
\begin{document}

\title{\ \\ \ \\ \ \\Einstein and Yang-Mills theories in hyperbolic form\\
without gauge-fixing}

\author{Andrew Abrahams\cite{AMA}, Arlen Anderson, Yvonne
Choquet-Bruhat\cite{CBadd}, and
James W. York Jr.}

\address{Department of Physics and Astronomy\\
         University of North Carolina\\
         Chapel Hill, NC 27599-3255
          }

\date{June 28, 1995}
\maketitle
\vspace{-11cm}
\hfill IFP-UNC-510

\hfill TAR-UNC-048

\hfill gr-qc/9506072
\vspace{10cm}
\begin{abstract}
The evolution of physical and gauge degrees of freedom in the Einstein
and Yang-Mills theories are separated in a gauge-invariant manner.
We show that the equations of motion of these theories can always be
written in flux-conservative
first-order symmetric hyperbolic form.  This dynamical form is ideal for
global analysis, analytic approximation methods such as gauge-invariant
perturbation theory, and numerical solution.
\end{abstract}
\newpage

One of the prevailing issues facing general relativity, indeed any gauge
theory, is the separation of physical from gauge degrees of freedom.  This
conceptual difficulty is encountered in generation of solutions of the field
equations, proofs of existence and uniqueness of solutions, and attempts
at quantization.  In this letter we present explicitly
{\it hyperbolic} forms\cite{CBY95} of the Einstein and Yang-Mills equations
of motion which clearly display the dynamics of these theories
{\it without fixing a gauge}.  (The constraint equations remain
elliptic.)  The basic strategy, which is applicable to any gauge theory, is
to take an additional time-derivative of the equations of motion, use the
constraint equations, and guarantee equivalence to the original
theory {\it via} appropriate choice of Cauchy data.  This completes the
program, begun by the French,
to cast general relativity in 3+1 form\cite{Lic,ADM,Yor79,CBY79,OMY}
by integrating it with the somewhat more recent efforts directed
towards finding a hyperbolic
formulation of general relativity\cite{CB52,FiM,CBR,Bea,VPE}.

Our hyperbolic formulation preserves complete spatial covariance
by means of an arbitrary shift vector. The standard 3+1 treatment
\cite{ADM,Yor79}, is gauge covariant in this sense but {\it not}
hyperbolic.  Our formulation does require a condition on
the time slicing to deal with the time-reparametrization invariance of
the theory.

A hyperbolic formulation of general relativity is valuable
for many applications.  The study of analytic approximations can be given
a rigorous foundation.  Gauge-invariant perturbation theory\cite{Mon}
arises naturally as a perturbative reduction of the new equations\cite{AAY}.
Problems in global analysis, regarding the existence
and uniqueness of solutions\cite{MoE,ChK}, take on a new light when
viewed with the powerful tool of hyperbolic theory\cite{Ler,CBP,FiM}.
Insights into quantum gravity and the problem of time
seem likely, given an understanding of the precise role of time slicing
necessitated by hyperbolicity.

For numerical relativity, the importance of a hyperbolic formulation
cannot be overstated.  There are many algorithms for solving the hyperbolic
equations fluid dynamics which can now be applied to general relativity.
More fundamentally, the isolation of physical from gauge effects
means that a numerically generated spacetime can have closer connection
with a desired astrophysical scenario.  In the ``Grand Challenge''
effort to solve Einstein's equations for the inspiral and coalescence of
compact binaries---a process expected to be observable in gravitational
radiation by LIGO and other detectors---one of the difficult
problems is the treatment of the horizon of the
black hole:  gauge degrees of freedom can propagate faster than
light and can thus escape from the black hole.  In a hyperbolic
formulation whose only non-zero speed of propagation is that of light, the
horizon again becomes a natural physical boundary.
The hyperbolic formulation is also ideal for the treatment
of gravitational radiation in numerically generated spacetimes
as it makes manifest a split between background and propagating radiation
which has long been assumed in approximate calculation schemes \cite{Kip},
in the extraction of gravitational radiation waveforms at finite radius,
and in the imposition of outgoing wave boundary conditions \cite{Abr}.

We first demonstrate the procedure in the simpler context
of Yang-Mills field theory in flat spacetime (cf. \cite{CBP}).
The Yang-Mills field strength $F^{a}\vphantom{|}_{\mu\nu}$ is given in
terms of the vector gauge potential $A^{a}\vphantom{|}_\mu$ by
\begin{equation}
\label{dA}
F^{a}\vphantom{|}_{\mu\nu}= \partial_\mu A^{a}\vphantom{|}_{\nu}
-\partial_\nu A^{a}\vphantom{|}_{\mu}
+f^{a}\vphantom{|}_{bc}A^{b}\vphantom{|}_\mu A^{c}\vphantom{|}_\nu,
\end{equation}
where $\partial_\mu$ indicates ordinary partial differentiation in the
$x^\mu$ direction in a Minkowski coordinate frame and
$f^{a}\vphantom{|}_{bc}$ are the structure constants of the Yang-Mills
gauge group $G$.  With $D_\mu$ indicating a gauge
covariant derivative, the Yang-Mills field equations in the absence of sources
consist of three equations of motion (for each gauge index value)
\begin{equation}
\label{YMeom}
D^0 F^{a}\vphantom{|}_{i0}+D^j F^{a}\vphantom{|}_{ij}=0
\end{equation}
and a constraint
\begin{equation}
\label{YMcons}
D^j F^{a}\vphantom{|}_{0j}=0.
\end{equation}
The Bianchi identity is
\begin{equation}
\label{YMbianchi}
0=D_\lambda F^{a}\vphantom{|}_{\mu\nu} +D_\mu F^{a}\vphantom{|}_{\nu\lambda}
+D_\nu F^{a}\vphantom{|}_{\lambda\mu}.
\end{equation}
This is identically satisfied given the definition of
$F^{a}\vphantom{|}_{\mu\nu}$ and does not need to be separately imposed.

A hyperbolic wave equation for $F^{a}\vphantom{|}_{i0}$ is obtained by taking
a covariant time derivative of (\ref{YMeom}) and subtracting a spatial
gradient of (\ref{YMcons}):
\begin{equation}
D_0 D^0 F^{a}\vphantom{|}_{i0}+D_0 D^j F^{a}\vphantom{|}_{ij}-
D_i D^j F^{a}\vphantom{|}_{0j}=0.
\end{equation}
Interchanging the order of covariant differentiations produces (gauge)
curvature terms which may be combined using the antisymmetry of the structure
constants.  A covariant divergence of the Bianchi identity is then used
to give the non-linear wave equation
\begin{equation}
\label{YMweq}
D^\mu D_\mu F^{a}\vphantom{|}_{i0}
+2f^{a}\vphantom{|}_{bc}F^{b}\vphantom{|}_{ij}F^{cj}\vphantom{|}_{0}=0.
\end{equation}

The full second-order system of equations consists of the wave equation
(\ref{YMweq}), the constraint (\ref{YMcons}), and the definition of
$F^{a}\vphantom{|}_{\mu\nu}$ in (\ref{dA}). [Combining the definition
of $F^{a}\vphantom{|}_{i0}$ in (\ref{dA}) with the wave equation
(\ref{YMweq}) would produce a third-order hyperbolic equation, with
principal part $\Box \partial/\partial t$, for $A^{a}\vphantom{|}_i$.]
This system is hyperbolic with elliptic conditions for
initial data, and its solution is unique once Cauchy data have been
specified on an initial spacelike hypersurface.  The Cauchy data consist of
an arbitrary gauge potential $A^{a}\vphantom{|}_0$, the pair
$A^{a}\vphantom{|}_i$ and $F^{a}\vphantom{|}_{i0}$ consistent with the
constraint (\ref{YMcons}), and $D^0 F^{a}\vphantom{|}_{i0}$ such that
the Yang-Mills equation of motion (\ref{YMeom}) holds on the initial slice.
With these data, one can prove the hyperbolic system is equivalent to the
original Yang-Mills equations, yet no gauge-fixing condition has been
imposed\cite{Gribov}.

The equations of motion (\ref{YMweq}) and the definition of
$F^{a}\vphantom{|}_{i0}$ can be put in flux-conservative
first-order symmetric hyperbolic form.  The magnetic part of the
equations, implicit in the Bianchi identity, must
now be used explicitly.  Introducing the derivatives of the field
strength
$G_{\lambda}\vphantom{|}^{a}\vphantom{|}_{\mu\nu}
=D_{\lambda} F^{a}\vphantom{|}_{\mu\nu}$ as new
variables, one finds
\begin{eqnarray}
\label{YM1a}
D^0 G_0\vphantom{|}^{a}\vphantom{|}_{\mu\nu}
+ D^k G_k\vphantom{|}^{a}\vphantom{|}_{\mu\nu}&=&
-2 f^{a}\vphantom{|}_{bc} F^{b}\vphantom{|}_{\mu}\vphantom{|}^{\lambda}
F^{c}\vphantom{|}_{\lambda\nu}, \\
\label{YM1b}
D^0 G_k\vphantom{|}^{a}\vphantom{|}_{\mu\nu}
-D_k G\vphantom{|}^{0a}\vphantom{|}_{\mu\nu}&=&
-f^{a}_{bc}F^{b}\vphantom{|}_{k}\vphantom{|}^{0}
F^{c}\vphantom{|}_{\mu\nu}.
\end{eqnarray}
The unknowns of the first-order hyperbolic system are $A^{a}\vphantom{|}_i$,
$F^{a}\vphantom{|}_{\mu\nu}$, and
$G_\lambda\vphantom{|}^{a}\vphantom{|}_{\mu\nu}$, and the
equations consist of the definitions of $F^{a}\vphantom{|}_{i0}$ and
$G_0\vphantom{|}^{a}\vphantom{|}_{\mu\nu}$, (\ref{YM1a}) and (\ref{YM1b}).

 From the first-order form, one sees that $A^{a}\vphantom{|}_j$,
and $F^{a}\vphantom{|}_{\mu\nu}$ propagate with speed zero:
that is, they are simply ``dragged along'' the time axis during the
evolution.  It is only the derivatives of the
field strength that propagate with the speed of light ($c=1$).
It must be emphasized that $A^{a}\vphantom{|}_{0}$ is {\it not}
a characteristic field and that only the fields which propagate with
non-zero speed are gauge covariant.  A gauge must be chosen to specify
$A^{a}\vphantom{|}_j$, but no gauge-fixing condition is required for
hyperbolicity.

A hyperbolic formulation for general relativity can be found by
a similar procedure\cite{CBY95}.  (Cf. \cite{CBR} where complete spatial
gauge covariance is not present because of the choice of a zero shift
vector.) Consider a globally hyperbolic manifold of topology
$\Sigma\times R$ with the metric
\begin{equation}
ds^2 = -N^2 dt^2 +g_{ij} (dx^i +\beta^i dt) (dx^j +\beta^j dt),
\end{equation}
where $N$ is the lapse and $\beta^i$ is the shift.  Introduce the
non-coordinate co-frame,
\begin{equation}
\theta^0 = dt, \quad
\theta^i = dx^i+\beta^i dt.
\end{equation}
with corresponding dual (convective) derivatives
\begin{equation}
\partial_0 = \partial/\partial t -\beta^i \partial/\partial x^i, \quad
\partial_i = \partial/\partial x^i.
\end{equation}
Note that $[\partial_0,\partial_i]=(\partial_i \beta^k)\partial_k
=C_{0i}\vphantom{|}^{k}\partial_k$, where the
$C$'s are the structure functions of the co-frame,
$d\theta^{\alpha}= -{1\over 2}C_{\beta\gamma}\vphantom{|}^{\alpha}
\theta^{\beta}\wedge \theta^{\gamma}$.

The natural time derivative for evolution is\cite{Yor79}
\begin{equation}
\hat\partial_0=\partial_0 +\beta^k \partial_k -{\cal L}_{\beta}
=\partial/\partial t -{\cal L}_{\beta},
\end{equation}
where ${\cal L}_{\beta}$ is the Lie derivative in a time slice $\Sigma$ along
the shift vector.  In combination with the lapse as
$N^{-1}\hat\partial_0$, this is the derivative with respect to proper time
along the normal to $\Sigma$, and it always lies inside the light cone,
in contrast to $\partial/\partial t$.  It has the useful property that
$[\hat\partial_0,\partial_i]=0$. The extrinsic curvature $K_{ij}$ of
$\Sigma$ is given by
\begin{equation}
\label{dg}
\hat\partial_0 g_{ij}= -2 N K_{ij}.
\end{equation}

One employs a procedure
parallel to that used in Yang-Mills theory.  The spatial metric $g_{ij}$
is analogous to $A^{a}\vphantom{|}_{i}$, the shift $\beta^k$ to
$A^{a}\vphantom{|}_{0}$, and the extrinsic curvature $K_{ij}$ of
$\Sigma$ to $F^{a}\vphantom{|}_{i0}$.
The lapse $N$ is a new feature present in time-reparametrization invariant
theories.

In four-dimensions, Einstein's theory, $R_{\mu\nu}=8\pi G(T_{\mu\nu}
-{1\over 2}g_{\mu\nu} T^{\lambda}\vphantom{|}_{\lambda})$,
leads to six equations of motion from $R_{ij}$, three ``momentum
constraints'' from $R_{0i}$,
and the Hamiltonian constraint from
$G^{0}\vphantom{|}_{0}={1\over 2}(R^{0}\vphantom{|}_{0}-
R^{k}\vphantom{|}_{k})$.  The hyperbolic form of Einstein's theory is
obtained by taking a time derivative of the equations of motion and
subtracting spatial gradients of the momentum constraints,
\begin{equation}
\label{omega}
\hat\partial_0 R_{ij} -\bar\nabla_{i} R_{0j}-\bar\nabla_{j}
R_{i0}=\Omega_{ij}.
\end{equation}
(Barred quantities are defined in the hypersurface $\Sigma$.)

Expressing (\ref{omega}) in a 3+1 decomposition, one finds
\begin{equation}
\label{boxK}
\Omega_{ij}=N\Box K_{ij} + J_{ij} +S_{ij},
\end{equation}
where $\Box=-N^{-1}\hat\partial_0 N^{-1}\hat\partial_0
+\bar\nabla^{k} \bar\nabla_{k}$ is the physical wave operator for
arbitrary $\beta^k$.
If we denote the trace of the extrinsic curvature by
$H=K^{k}\vphantom{|}_{k}$, then
\begin{eqnarray}
\label{Jij}
J_{ij}&=& \hat\partial_0 (H K_{ij} - 2 K_{i}\vphantom{|}^{k} K_{jk})
 +(N^{-2}\hat\partial_0 N+ H)\bar\nabla_i \bar\nabla_j N \nonumber \\
&&\hspace{-0.75cm}
-2N^{-1}(\bar\nabla_k N) \bar\nabla_{(i}(N K^{k}\vphantom{|}_{j)})
+3 (\bar\nabla^k N) \bar\nabla_k K_{ij} \\
&&\hspace{-0.75cm} +N^{-1}K_{ij} \bar\nabla^k (N\bar\nabla_k N)
-2 \bar\nabla_{(i}(K_{j)}\vphantom{|}^{k}\bar\nabla_k N)
-N\bar\nabla_i\bar\nabla_j H
\nonumber \\
&&\hspace{-0.75cm}
+N^{-1}\bar\nabla_i\bar\nabla_j(N^2 H)
 -2N K^{k}\vphantom{|}_{(i}\bar R_{j)k}
-2N \bar R_{kijm}K^{km}. \nonumber
\end{eqnarray}
[where $M_{(ij)}={1\over 2} (M_{ij}+M_{ji})$] and
\begin{equation}
S_{ij}=-N^{-1}\bar\nabla_i \bar\nabla_j(\hat\partial_0 N +N^2 H).
\end{equation}

For $\Omega_{ij}$ to produce a wave equation, $S_{ij}$  must be equal
to a functional involving fewer than second derivatives of
$K_{ij}$.  This can apparently be accomplished in a number of ways and
constitutes the imposition of a slicing condition on the spacetime.
It is necessary to show that the slicing condition can be imposed
without spoiling the hyperbolic nature of the evolution system.

A clear and simple slicing condition is the harmonic condition (cf.
\cite{CBR} when $\beta^k=0$)
\begin{equation}
\label{d0N}
\hat\partial_0 N +N^2 H=0.
\end{equation}
(This can easily be generalized by adding an ordinary well behaved function
$f(t,x)$ to the right hand side.)
Imposing (\ref{d0N}) for all time amounts to imposing an equation of motion
for $N$.  The complete system of equations of motion is now the wave
equation (\ref{boxK}) for $K_{ij}$, the harmonic slicing condition
(\ref{d0N}), and the definition (\ref{dg}) of the extrinsic curvature.
The Cauchy data for the full system, to be given on an initial slice $\Sigma$,
are $g_{ij}$ and $K_{ij}$ consistent with the Hamiltonian and momentum
constraints, the lapse $N$, and $\hat\partial_0 K_{ij}$ such that the
Einstein equations of motion hold on the initial slice.  Using the doubly
contracted Bianchi identity, one can prove\cite{CBY95}  that, with these
initial data, the hyperbolic system is fully equivalent to Einstein's
theory.

Another useful class of slicing conditions arises from choosing
$H$ to be a known function of spacetime $h(t,x)$.
In this case, the lapse function $N$ is determined by
solution of the time-dependent elliptic problem:
\begin{equation}
\hat \partial_0 h= - \bar\nabla^k\bar\nabla_k N
+ N (\bar R + H^2 -g^{ij}R_{ij}).
\end{equation}
In this scheme, the Cauchy data are simply $g_{ij}$ and $K_{ij}$
satisfying the constraints and
$\hat\partial_0 K_{ij}$ from the usual evolution equation.
The $g_{ij}, K_{ij}$ system is still hyperbolic, but the full set of
equations is now mixed hyperbolic-elliptic.  The shift vector can still be
specified arbitrarily.  Proof of a unique solution proceeds by an iterative
method, and equivalence with the usual form of Einstein's equations
again employs the twice-contracted Bianchi identity\cite{CBY95}.

The harmonic condition is consistent with the natural slicings of
stationary spacetimes.  For example, suppose one has a spacetime
with a timelike Killing vector and a spacelike Killing vector
proportional to the shift vector: $\beta^i = f \xi^i$.  In this
case the evolution of the 3-metric gives $\hat \partial_0 g_{ij}=
-2 \xi_{(i} \partial_{j)} f$ so $H=N^{-1} \xi^l\partial_l f$.
Eliminating $H$ with the harmonic slicing condition
yields $\xi^l \partial_l (f/N)=0$, which is clearly true for the Kerr geometry
in Boyer-Lindquist
coordinates.  It is possible and useful in perturbation theory  to have
the advantages of a specified $H$ and the harmonic slicing condition
by choosing the shift vector suitably.  However, choosing a shift vector in
this particular way seems undesirable for numerical solution of the
full field-equations because
spacetimes evolved in this fashion will tend to
develop coordinate singularities as in the stationary spacetimes mentioned
above.

In the vacuum case, if we introduce, besides $N$, $g_{ij}$ and $K_{ij}$,
new variables $a_i=N^{-1}\bar\nabla_i N$---the acceleration of the local
Eulerian observers (those at rest in the time slices)---its derivatives
$a_{0i}=N^{-1}\hat\partial_0 a_i$ and
$a_{ji}=\bar\nabla_j a_i=a_{ij}$, as well as the derivatives of
the extrinsic curvature
\begin{equation}
\label{d0K}
\hat\partial_0 K_{ij}=N L_{ij}
\end{equation}
and
$M_{kij}=\bar\nabla_k K_{ij}$,  one can cast the
equations (\ref{dg}), (\ref{boxK}), (\ref{d0N}) into complete
flux-conservative first-order symmetric hyperbolic form\cite{CBY95}.
The unknowns of the first-order system are $g_{ij}$, $N$, $K_{ij}$
$L_{ij}$, $M_{kij}$, $a_i$, $a_{ji}$ and $a_{0i}$, and the equations of
the first order system are  (\ref{dg}), (\ref{d0N}), (\ref{d0K}) and
\begin{equation}
\hat\partial_0 L_{ij} - N \bar\nabla^{k} M_{kij} = N( H L_{ij} -J_{ij}),
\end{equation}
\begin{eqnarray}
\hat\partial_0 M_{kij} -N \bar\nabla_{k} L_{ij} &=&
N[ a_k L_{ij} + 2M_{k(i}\vphantom{|}^{m} K_{j)m}\\
&&\hspace{-2.5cm} + 2K_{m(i} M_{j)k}\vphantom{|}^{m}
-2K_{m(i} M^{m}\vphantom{|}_{j)k}  \nonumber\\
&&\hspace{-2.5cm} + 2K_{m(i} ( K^{m}\vphantom{|}_{j)}a_k
+ a_{j)} K^{m}\vphantom{|}_{k} -a^{m} K_{j)k})] , \nonumber
\end{eqnarray}
\begin{equation}
\hat\partial_0 a_i =-N(H a_i + M_{ik}\vphantom{|}^{k}),
\end{equation}
\begin{eqnarray}
\hat\partial_0 a_{ji} -N \bar\nabla_{j} a_{0i}&=& Na_{k}
[2 M_{(ij)}\vphantom{|}^{k} -M^{k}\vphantom{|}_{ij} \\
&&\hspace{-2cm}+2 a_{(i} K_{j)}\vphantom{|}^{k} - a^{k} K_{ij}]
 + N a_j a_{0i}, \nonumber
\end{eqnarray}
\begin{eqnarray}
\hat\partial_0 a_{0i} - N\bar\nabla^{k} a_{ki} &=&
N [-\bar R^{k}\vphantom{|}_{i} a_{k} +a_{i}( H^2 -2 K_{kl}K^{kl} \\
&&\hspace{-3cm}+2 a^k a_k +2 a^{k}\vphantom{|}_{k})
+2 a_k a^{k}\vphantom{|}_{i} + H M_{ik}\vphantom{|}^{k}
-2K^{kl} M_{ikl}],  \nonumber
\end{eqnarray}
where $J_{ij}$ can be found from (\ref{Jij}).  Notice
that the shift is {\it not} one of the characteristic fields.  The form of
the first-order system is independent of the choice of $\beta^k$, though
it must be specified for solutions. To
complete the reduction to first-order form,
the 3-dimensional Riemann curvature appearing in $J_{ij}$
is expressed in terms of the 3-dimensional Ricci curvature using
\begin{equation}
\label{barR}
\bar R_{mijk}= 2 g_{m[j}\bar R_{k]i} + 2g_{i[k}\bar R_{j]m}+
\bar R g_{m[k} g_{j]i},
\end{equation}
which in turn is eliminated by substituting
\begin{equation}
\bar R_{ij}= R_{ij} +L_{ij} -H K_{ij} +2 K_{ik}K^{k}\vphantom{|}_j +
a_i a_j+ a_{ji}.
\end{equation}
The four-dimensional Ricci curvature is then eliminated using the
Einstein equations.

Note that in spatial dimensions greater than three, the
expression for $\bar R_{mijk}$ involves the Weyl tensor, which
cannot be eliminated using the Einstein equations.
The reduction to first order form in these variables is thus
blocked.

One sees that $g_{ij}$, $K_{ij}$, $N$, and $a_i$ all propagate with zero
speed with respect to the Eulerian observers: they are dragged along
the normal to the foliation by the evolution.
Only the derivatives of the extrinsic curvature and the derivatives of the
acceleration propagate with the speed of light.  These represent
time-dependent tidal forces and can be used to form the components of the
spacetime Riemann tensor.  The only propagating degrees of freedom then
are curvatures, as one would expect physically.  Equivalently, in the
second order system, the wave equation for $K_{ij}$
can be viewed as defining the notion of radiation as distinctly as
possible in the context of a nonlinear, curved space field theory.
The inherent separation of the evolution of the spatial-metric and extrinsic
curvature is a natural starting point for a formal expansion
scheme which is gauge-invariant at each order.

When linearized around a static background 3-metric,
the evolution equations for $g_{ij}$ and
the wave equation for $K_{ij}$ decouple.   For instance, with
a flat space background, the simple wave equation $\Box K_{ij}=0$ for
the extrinsic curvature is obtained (assuming harmonic slicing).
The evolution equation for the 3-metric
contains no new information about the
dynamical degrees of freedom.  Similarly, one can linearize about
static or stationary black hole backgrounds, for instance Schwarzschild
or Kerr.   In the Schwarzschild case, (\ref{boxK}) reduces directly
to scalar wave equations for the even
and odd-parity radiation modes.  (The 3-metric evolution
equation is again irrelevant.)   Taking an additional time
derivative of the scalar equations, one recovers the standard results
of gauge-invariant perturbation theory~\cite{Mon}.
The pair of scalar wave equations
obtained for each ($\ell$, $m$) multipole combination
can be matched directly onto a numerically generated interior
solution and provides both a gauge-invariant
radiation extraction method and a clean prescription for
outer boundary conditions (including backscatter of waves).
By refining the assumed exterior background spacetime, arbitrary
amounts of physical detail can be incorporated by this general
method.

The reasoning we have applied in this paper can be applied to general
relativity coupled to other fields with well-posed Cauchy problems, as
well as to generally covariant and gauge theories in the broad sense.  One
sees that the procedure of taking time derivatives and adding further
variables can be continued to build a ``tower'' of equations which,
provided suitable initial conditions are given, is equivalent to the
original theory. Nothing fundamental is gained by
going beyond the stage at which gauge-invariant equations of motion are
obtained, as in this paper, but we find the equations that propagate
the spacetime Riemann tensor components directly aesthetically
appealing.  By achieving a hyperbolic formulation of a gauge theory
without gauge-fixing, one has manifestly physical propagation without
the encumbrance that comes from having to impose particular gauge conditions.
The physical structure of the theory is revealed with the full gauge
symmetry preserved.

A.A., A.A., and J.W.Y. were supported by National Science Foundation
grants PHY-9413207 and PHY 93-18152 (DARPA supplemented).


\begin{references}
\bibitem{Gribov}
This would deal with the Gribov ambiguity.
\end{references}
\end{document}